\documentclass[prl,superscriptaddress,showpacs,twocolumn]{revtex4}
\usepackage{graphicx}
\usepackage{graphicx,amsfonts}
\usepackage{epsfig,amsmath}
\usepackage{verbatim}

\begin{document}

\newcommand{\atanh}
{\operatorname{atanh}}
\newcommand{\ArcTan}
{\operatorname{ArcTan}}
\newcommand{\ArcCoth}
{\operatorname{ArcCoth}}
\newcommand{\Erf}
{\operatorname{Erf}}
\newcommand{\Erfi}
{\operatorname{Erfi}}
\newcommand{\Ei}
{\operatorname{Ei}}

\title{The longest excursion of stochastic processes in nonequilibrium
  systems}

\author{Claude Godr{\`e}che}
\affiliation{Institut de Physique Th\'eorique, IPhT, CEA Saclay, and URA 2306,
91191 Gif-sur-Yvette Cedex, France}
\author{Satya N. Majumdar}
\affiliation{Laboratoire de Physique Th\'eorique et Mod\`eles Statistiques
  (UMR du CNRS 8626), Universit\'e de Paris-Sud, 91405 Orsay Cedex, France}
\author{Gr{\'e}gory Schehr}
\affiliation{Laboratoire de Physique Th\'eorique (UMR du
 CNRS 8627), Universit\'e de Paris-Sud, 91405 Orsay Cedex, France}

\date{\today}

\begin{abstract}
We consider the excursions, {i.e.} the intervals between
consecutive zeros, of stochastic processes 
that arise in a variety of nonequilibrium systems
and study the temporal growth of the {\it longest}
one $l_{\max}(t)$ up to time $t$. 
For smooth processes, we find a universal linear growth $\langle l_{\max}(t) \rangle \simeq Q_{\infty}\,t 
$
with a model dependent amplitude $Q_\infty$. In contrast, for non-smooth processes with a
persistence exponent $\theta$, we show that $\langle l_{\max}(t) \rangle$ has
a linear growth if $\theta < \theta_c$ while $\langle l_{\max}(t) \rangle\sim
t^{1-\psi}$ if $\theta > \theta_c$. The amplitude $Q_{\infty}$ and the exponent $\psi$ are
novel quantities
associated to
nonequilibrium dynamics. These behaviors are obtained by exact
analytical  
calculations for renewal and
multiplicative processes and numerical simulations for other systems
such as the coarsening dynamics
in Ising model as well as the diffusion
equation with random initial conditions. 
\end{abstract}

\maketitle

{\it Introduction.} Nonequilibrium dynamics in many-body systems
keep offering new challenges despite several decades of research. 
An example of such a system, among others, is 
the Ising model undergoing phase ordering after
a rapid quench in temperature~\cite{bray}.
In such systems, the relevant stochastic process $X(t)$ that
represents, at a fixed point in space, the evolving spin in the Ising model (or e.g., the 
field in the diffusion equation)
is generically a complex one with nontrivial
history dependence. Traditional two-time correlation function $\langle X(t_1)X(t_2)\rangle$
is typically not sufficient to characterize the complex temporal history of such 
a process. One simple measure of this history dependence that has attracted much attention
in the recent past, both theoretically~\cite{claude_persistence, satya_review}
and experimentally~\cite{persist_exp}, is the persistence $p(t_1,t_2)$
defined as the probability that the process $X(t)$, adjusted to have zero mean,
has not changed sign in the interval $[t_1,t_2]$.
In several such nonequilibrium systems persistence $p(t_0,t)$, for
$t \gg t_0$, decays as a power law,
$p_0(t) = p(t_0,t \gg t_0) \sim t^{-\theta}$, with a nontrivial persistence 
exponent
$\theta$~\cite{satya_review}.
  
A stochastic process $X(t)$ (depicted
schematically
in Fig. 1) evidently does not change sign between two consecutive zero crossings.
The persistence $p_0(t)$ is simply related to the
probability distribution of time intervals (or excursions) between successive zeros
and is clearly one, but {\it not the only one}, possible characterization
of the history dependence of $X(t)$.
In this Letter we propose
an alternative yet simply measurable characteristic of the history 
of $X(t)$ via an {\it extreme} observable that elucidates, in a natural way, 
the important role played by {\it extreme value statistics} in such nonequilibrium 
systems. In particular, our results illustrate the {\it universal} features of 
extreme statistics in generic many-body nonequilibrium systems and provides, in addition,
interesting connections with the theory of records that attracted much interest recently in the 
context of
random walks~\cite{satya_ziff}, growing networks \cite{claude_jmluck} and
pinned elastic manifolds~\cite{pld}.   
\begin{figure}[t]
\includegraphics[width=\linewidth]{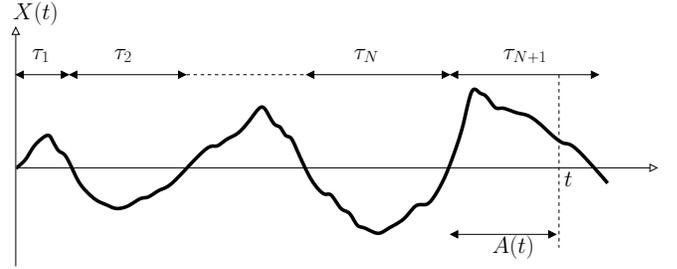}
\caption{Intervals between zero-crossings (excursions) of a stochastic
  process $X(t)$.}\label{Fig1}
\end{figure}

For a typical realization of a generic process $X(t)$ with
$N \equiv N(t)$ zeros in the {\em fixed} time interval $[0,t]$ (see~Fig.~\ref{Fig1}), let
$\{\tau_1,\tau_2,\cdots,\tau_N \}$ denote the interval lengths
between successive zeros and $A(t)$
denote the length (or age) of the last {\em unfinished} excursion.
Our proposed {\em extreme} observable is the length of the {\em longest}
excursion up to $t$
\begin{equation}
\label{def_lmax}
l_{{\rm max}}(t) = {\max} ( \tau_1, \tau_2, \cdots, \tau_N, A(t)) \;.
\end{equation} 

Let us first summarize our main results. We find rather rich universal late time behavior
of the average $\langle l_{\rm max}(t)\rangle$ for generic stochastic
processes $X(t)$ arising in nonequilibrium systems. 
Such processes typically belong to two broad classes~\cite{satya_review}: {\em smooth}
(i.e., with a finite density of zeros) and {\em non-smooth} (with
infinite density of zeros). While persistence typically decays algebraically, $p_0(t)\sim t^{-
\theta}$, irrespective of the smoothness of the process, 
$\langle l_{\rm max}(t)\rangle$, in contrast, turns out to be sensitive to the
smoothness of $X(t)$. 
For smooth processes, $\langle l_{\rm max}(t)\rangle$ always grows linearly with time
\begin{eqnarray}\label{theta_le1}
\langle l_{\max}(t) \rangle \simeq Q_{\infty} \; t  \;,
\end{eqnarray}
albeit with a model dependent prefactor $Q_{\infty} > 0$. In contrast, for
non-smooth processes, it grows either as in~(\ref{theta_le1}), or as 
\begin{eqnarray}\label{theta_ge1}
\langle l_{\max}(t) \rangle \sim t^{1-\psi} \;,
\end{eqnarray}
where the exponent $0<\psi<1$ (sublinear growth),
depending on whether the associated persistence exponent $\theta$ of the process 
is less ($\theta<\theta_c$) or greater ($\theta>\theta_c$) than a critical
value $\theta_c$.  
We establish these behaviors via exact analytical
calculations for two simple models, one corresponding to each class:
multiplicative (smooth) and renewal (non-smooth) processes. In addition, we
perform extensive numerical simulations in
a variety of nonequilibrium systems, including the diffusion
equation with random initial conditions and coarsening dynamics of the
Ising model both below and exactly at the critical temperature $T=T_c$. 
For the latter case, our results (\ref{theta_le1}, \ref{theta_ge1}) 
provide new universal quantities associated to nonequilibrium
critical dynamics. 

The full knowledge of the distribution of $l_{\max}$~\cite{us_prep} allows in particular the computation of its average.
However, here we compute $\langle l_{\max}(t) \rangle$ by using the relationship
\begin{eqnarray}\label{rel_l_q}
d\langle l_{\max}(t)\rangle/dt = Q(t)\;,
\end{eqnarray}
where 
$Q(t)$
denotes the probability that the last excursion in $[0,t]$,
$A(t)$ in~Fig.~\ref{Fig1}, is the longest one
\begin{eqnarray} \label{def_Q}
Q(t) = {\rm Prob}[l_{\rm max}(t) = A(t)] \;.
\end{eqnarray}
Thus $Q(t)$ is the rate at which the ``record'' length of
an excursion is broken at time $t$.
Indeed,
if the total interval increases from $t$ to $t+dt$, the random variable
$l_{\rm max}(t)$ either increases by $dt$ (if the last excursion
happens to be the longest one and the probability for this event is $Q(t)$)
or stays the same (with probability $1-Q(t)$). Taking average gives (\ref{rel_l_q}).
Henceforth we focus on $Q(t)$, rather than $\langle l_{\max}(t) \rangle$ directly.

{\it Renewal processes}. Let $X(t)$ be a renewal process with successive interval lengths 
$\tau_i$'s 
being independent random variables, each distributed according to a L\'evy law
with parameter $\theta$, $\rho(\tau) \sim \tau^{-1-\theta}$, for large $\tau$~\cite{renewal_gl}. The 
persistence is simply 
$p_0(t) = \int_{t}^\infty\, d\tau \rho(\tau) \sim t^{-\theta}$. 
The joint distribution $q_N(\tau_1,\tau_2, \cdots,
\tau_N,A(t);t)$ of the intervals depicted in~Fig.~\ref{Fig1} is then 
\begin{eqnarray}\label{blocks}
&&q_N(\tau_1,\cdots,\tau_N,A(t);t) = \rho(\tau_1) \rho(\tau_2) \cdots
\rho(\tau_N) p_0(A(t)) \nonumber \\
&&\times \;\delta(\tau_1 + \tau_2 + \cdots + \tau_N + A(t) - t) \;,
\end{eqnarray}
where the $\delta$ function ensures that the total interval length is $t$.
For the last interval to be the longest, the others must be shorter
than it implying that $Q(t)$ in
Eq.~(\ref{def_Q}) is 
\begin{equation}\label{expr_gen}
Q(t) = \sum_{N=0}^\infty \int_0^\infty db \int_0^b d\tau_1...
\int_0^b d\tau_N  q_N(\tau_1,\cdots,\tau_N,b;t) \;.
\end{equation}
Taking Laplace transform of Eq.~(\ref{expr_gen}) gives
a simple form for 
$\hat Q(s) = \int_0^\infty \, dt e^{-st} Q(t)$~:
\begin{eqnarray}
\hat Q(s) = \int_0^\infty \, db \frac{p_0(b) e^{-s b}}{1 - \int_0^b d\tau
\rho(\tau) e^{-s\tau}} \;.
\end{eqnarray}
Using $\rho(\tau) = -p_0'(\tau)$ in an integration
by part, followed by change of variables $b = x/s$ and $\tau =
y/s$, lead to an expression convenient for late time asymptotic analysis
\begin{eqnarray}\label{expr_q_simpl}
\hat Q(s) = \frac{1}{s}\int_0^\infty dx \frac{p_0(x/s) e^{-x}}{p_0(x/s)e^{-x}
  + \int_0^x \, dy \, p_0(y/s) e^{-y}} .
\end{eqnarray}

\begin{figure}[h]
\includegraphics[angle=0,width=\linewidth]{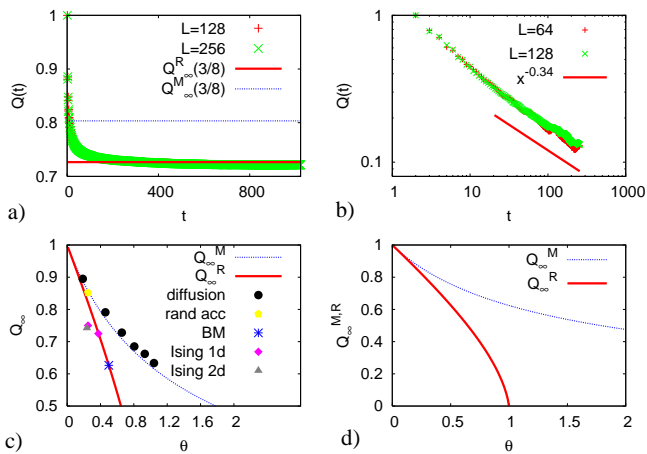}
\caption{{\bf a):} $Q(t)$ as a function of $t$ for the {\it local}
  magnetization in the Ising chain evolving with Glauber dynamics at
  $T=0$. {\bf b):} $Q(t)$ as a function of $t$ on a log-log scale for the
  magnetization of a line 
  in the $2d$-Ising model at $T_c$ evolving with Glauber dynamics
  starting from a fully magnetized state. {\bf c):} $Q_{\infty}$ as a function
  of $\theta$: the lines correspond respectively to $Q^R_{\infty}(\theta)$ (solid) and 
$Q^M_{\infty}(\theta)$ (dotted) and the points correspond to the numerical
values obtained for different nonequilibrium systems. The two values for `Ising 1d' correspond to the local magnetization ($\theta=3/8$) and to the global one ($\theta=1/4$). 
{\bf d)}: Same plot as in c) on a larger scale.}\label{Fig2} 
\end{figure}

For $\theta < 1$, one can take the limit $s \to 0$ directly in Eq.~(\ref{expr_q_simpl}), 
using $p_0(t) \sim
(t_0/t)^{\theta}$ for large $t$ with 
some non-universal microscopic time scale $t_0$. Interestingly, $t_0$ cancels between the numerator 
and the 
denominator in Eq.~(\ref{expr_q_simpl}), yielding $\hat Q(s)~\sim~Q^R_{\infty}/s$ and thus
$Q(t)\to Q_\infty^R $ for large $t$ (the superscript $R$ refers to {\it renewal} process), 
as announced in Eq.~(\ref{theta_le1})
with a universal constant that depends only on $\theta$ (and not on other details) 
\begin{eqnarray}\label{approx_q}
Q_\infty^R \equiv Q^R_{\infty}(\theta) = \int_{0}^\infty \frac{dx}{ 1 +
  x^{\theta} e^x 
  \int_0^x \, dy \, y^{-\theta} e^{-y}} \, .
\end{eqnarray}
A special case of this general result, $\theta=1/2$, corresponds
to Brownian motion if one considers only ``large'' excursions, {\it
  i.e.} 
$\tau_i$'s in~Fig.~\ref{Fig1} larger than some cut-off 
$\tau_\epsilon$. This recovers in a simple way the result $Q^R_{\infty}(1/2) = 0.626508...$,
derived previously by mathematicians~\cite{pitman_yor} using rather complicated, albeit rigorous,
method. Note that $Q^R_{\infty}(\theta)$ in Eq.~(\ref{approx_q}) vanishes as $\theta\to 1$.
A plot of $Q^R_{\infty}(\theta)$ vs. $\theta$ is shown in~Fig.~\ref{Fig2}d.  

In contrast, for $\theta>1$, the naive substitution of $p_0(t=y/s)\sim (t_0/t)^{\theta}$ 
in the integral in the 
denominator 
of Eq. 
(\ref{expr_q_simpl}) is problematic since the integral
diverges. Instead a careful analysis shows that 
$\int_0^x \, dy \, p_0(y/s) e^{-y} \propto \langle \tau \rangle\, s$
as $s \to 0$ where $\langle \tau \rangle = \int_0^\infty \, d\tau \, \tau
\rho(\tau)$. This yields, after simple algebra, $\hat Q(s)
\sim s^{-1/\theta}$ and thus $Q(t) \sim t^{-1+1/\theta}$, as
announced in Eq.~(\ref{theta_ge1}), with
\begin{eqnarray}\label{approx_theta}
\psi = 1 - 1/\theta \;.
\end{eqnarray}
Thus for renewal processes the change of behavior of $\langle l_{\rm max}(t)\rangle$
happens at $\theta=\theta_c=1$. Qualitatively this transition can be
understood by simple scaling arguments combining extreme value statistics with
the behavior of the sum of independent L\'evy
variables~\cite{us_prep}.
The fact that the asymptotics of $\langle l_{\rm max}(t)\rangle$ for Brownian motion, a highly 
``non-smooth'' process with infinite density of zero-crossings, corresponds to
a special case ($\theta=1/2$) of the renewal process suggests
that the latter might qualitatively lead to a good approximation of
$\langle l_{\rm max}(t)\rangle$ for other non-smooth
processes such as the coarsening dynamics of the Ising model, 
and leads us to hypothesise that this
change of asymptotic behavior of $\langle l_{\rm max}(t)\rangle$ at a certain $\theta_c$
might be generic for non-smooth processes. 
Such an approximation of the phase ordering of the Ising model by a renewal process is
also useful for other observables~\cite{balda}.
However, for generic non-smooth
processes with $\theta>\theta_c$, the 
scaling
relation $\psi=1-1/\theta$ obtained under renewal approximation is in
general not valid and $\psi$ seems to be a new exponent.
Numerical results indeed support 
this hypothesis.

{\it Multiplicative processes.} A process $X(t)$ is multiplicative if the locations
of its zeros $\{t_1,t_2,\cdots\}$ 
are such that the 
successive ratios $U_k=t_{k-1}/t_k$ are independent random variables, 
each distributed over $U\in [0,1]$ with density $\tilde \rho(U)$.
While the calculation of $Q(t)$ is difficult for arbitrary $\tilde \rho(U)$, it
turns out that for the special family of density parametrized
by $\theta$, $\tilde \rho(U) = \theta 
U^{\theta-1}$,
one can use recent results of Ref.~\cite{claude_jmluck} to show that 
$Q(t)\to Q_\infty^M$ 
(where $M$ refers to {\it multiplicative} process), leading
to a linear growth of $\langle l_{\rm max}(t)\rangle$ as in Eq.~(\ref{theta_le1}) 
with
\begin{eqnarray}\label{approx_r}
Q^M_{\infty} \equiv Q^M_{\infty}(\theta) = \int_0^\infty \, ds \, e^{-s -
  \theta E(s)} \;,
\end{eqnarray}
where $E(s) = \int_s^\infty \, dx \, e^{-x}/x$. In particular, for uniform distribution, 
$Q^M_{\infty}(1) = 0.624329...$, the
Golomb-Dickman constant that also describes the asymptotic linear growth of the
longest cycle of a random permutation~\cite{finch_book}. In~Fig.~\ref{Fig2}d we show a plot of $Q^M_{\infty}(\theta)$. At
variance with renewal processes, $\langle l_{\rm max}(t)\rangle \simeq Q^M_\infty(\theta)\, t$ 
for all~$\theta$.

To appreciate this result in a
more general context, we note that
a multiplicative process $X(t)$ is
non-stationary by construction. However, when plotted 
as a function of $T=\ln (t)$, the process becomes a stationary
renewal process in $T$ since the successive
intervals on the $T$ axis $T_{k}-T_{k-1}$ become statistically independent.
Similarly, it turns out that for many nonequilibrium processes (e.g, diffusion
equation with random initial condition), the original non-stationary process in real time $t$
becomes stationary in $T=\ln(t)$~\cite{satya_review} and then
the renewal approximation is precisely equivalent to the 
Independent Interval Approximation (IIA)~\cite{persist_diffusion},
known to be a very good one for smooth processes~\cite{satya_review}.
For such smooth processes then, multiplicative process is
a good approximation in real time $t$. Within the IIA, the
interval distribution in log-time $T$ decays as
$\sim \exp[-\theta T]$ for large $T$ where $\theta$ is the
associated persistence exponent. In real time $t$, this then
corresponds to a multiplicative process with $\tilde \rho(U)\sim \theta U^{\theta-1}$ for small $U$.
If one assumes further that this power law form of $\tilde \rho(U)$ holds over the full
range of $U\in [0,1]$ one arrives precisely at the model studied above with
the parameter $\theta$ being the persistence exponent.
Thus, the multiplicative process with $\tilde \rho(U)=\theta U^{\theta-1}$ 
seems to be qualitatively a good representative of generic smooth 
processes, leading to the hypothesis of the asymptotic linear growth
of 
$\langle l_{\rm max}(t)\rangle$ for such smooth processes. This hypothesis
is supported by numerical simulations.

{\it Numerical results.} We have computed $Q(t)$ for various processes
for which the persistence exponent $\theta$ is known either exactly
or numerically. Guided by our analytical results, we have considered both
non-smooth and smooth processes and the numerical results
are consistent with the two broad behaviors announced in Eqs.~(\ref{theta_le1}) and (\ref{theta_ge1}). 
In the first case, $Q(t) \to Q_{\infty}$
as in Eq.~(\ref{theta_le1}), as shown in Fig.~\ref{Fig2}a while in
the second case, $Q(t) \sim t^{-\psi}$ as in Eq.~(\ref{theta_ge1}), as
shown in Fig.~\ref{Fig2}b. 

As a prototype of non-smooth processes, we have studied the magnetization in
the coarsening dynamics of a $d$-dimensional
ferromagnetic Ising system of 
linear size $L$ consisting of $L^d$ spins $\sigma_i = \pm 1$, with
periodic boundary conditions (pbc). Starting from a random initial
condition, the spins evolve via
Glauber
dynamics with nearest neighbour Ising Hamiltonian 
$H_{\rm Ising} = - \sum_{\langle i,j \rangle} \sigma_i \sigma_{j}$.
Our results are summarized below:

$\bullet$ In $d=1$ and at zero temperature, 
we have computed $Q(t)$ for the 
local magnetization $X(t)=\sigma_i(t)$, for which $\theta = 3/8$~\cite{derrida_ising}. Fig.~\ref{Fig2}a shows a plot of $Q(t)$
vs. $t$ for $L=128$ and $256$. These data show
that $Q(t) \to Q_\infty$ 
with $Q_\infty
= 0.725(5)$, which is very close to the analytical value obtained for a renewal process 
in~Eq. (\ref{approx_q}) with $\theta=3/8$, for which $Q^R_{\infty}(3/8) =
0.726531..$ (while for a multiplicative process one has $Q_{\infty}^M(3/8) =
0.80338...$), see Fig.~\ref{Fig2}c.

$\bullet$ We obtained a similar behavior, { i.e.} $Q(t)
\to Q_\infty$ for the global magnetization $M(t) = L^{-d}
\sum_i \sigma_i(t)$ both in $d=1$ at $T=0$ (for which $\theta=1/4$
\cite{satya_critical}) and in $d=2$ at the critical point $T=T_c$ (for
which $\theta = 0.237(3)$~\cite{satya_critical}). These results are shown in
Fig.~\ref{Fig2}c. Note that for the global magnetization, the agreement between
the numerical  
value of $Q_\infty$ and the corresponding $Q_{\infty}^R(\theta)$ is only qualitative.
%
%

$\bullet$ The exact solution obtained for renewal processes shows that if
$\theta$ is large enough (in that case $\theta > 1$), $Q(t)$ decays to 
zero as a power law $Q(t) \sim t^{-\psi}$ (\ref{theta_ge1},
\ref{approx_theta}). Recently, it was shown that for critical dynamics
of Ising systems  
starting from a completely ordered state, the persistence exponent
associated to $X(t) = M(t) - \langle M(t) \rangle$ where $M(t)$ is the
global magnetization can be large, for instance $\theta = 1.7(1)$ in
$2d$~\cite{raja_epl}. Unfortunately, the numerical computation of
$Q(t)$ is quite difficult in that case because the 
exponent $\psi$ is seemingly positive but very small. Alternatively,
starting from a fully magnetized state, one can instead consider, as
in Ref.~\cite{satya_manifold}, the process $X(t) = M_l(t) - \langle
M(t) \rangle$ where $M_l(t)$ is the magnetization of a {\it line}, for
which the persistence exponent is even larger $\theta \simeq 3.3$
\cite{raja_tbp}. 
In Fig.~\ref{Fig2}b,
we show a plot of $Q(t)$ for 
this process for two different system sizes $L = 64, 128$. This plot
is compatible with a power law decay $Q(t) \sim t^{-\psi}$ 
with $\psi = 0.34(1)$, which is actually rather far from
the value obtained for a renewal process
in~Eq.~(\ref{approx_theta}) which gives $\psi = 1-1/\theta \simeq 0.7$. 

These results for non-smooth processes in coarsening dynamics, 
summarized in Fig.~\ref{Fig2}c, are qualitatively and in some cases even
quantitatively (see 
Fig.~\ref{Fig2}a) in agreement with the results for
renewal processes in Eqs.~(\ref{approx_q},~\ref{approx_theta}). 

As a prototype of smooth processes, we have studied the diffusing
field $\varphi(\mathbf{x},t)$ evolving according to the heat equation 
$\partial_t \varphi(\mathbf{x},t) = \nabla^2 
\varphi(\mathbf{x},t)$ with pbc in dimension $d$ starting from random initial condition
$\langle \varphi(\mathbf{x},t=0) 
\varphi(\mathbf{x'},t=0) \rangle = 
\delta^d(\mathbf{x}-\mathbf{x'})$. It is known that the
persistence exponent $\theta \equiv \theta(d)$ 
associated to the diffusing field at the origin $X(t)=\varphi(\mathbf{x=0},t)$ depends
continuously on $d$~\cite{persist_diffusion} and in particular $\theta(d=46)
\simeq 1$ 
\cite{newman}. The probability $Q(t)$ can be easily computed
numerically in any dimension 
$d$ by solving the heat equation and noticing that the field at the
origin $\varphi({\mathbf 0},t)$ can be simply written, for a large
system size, as $\varphi({\mathbf 0},t) \sim \int_0^\infty \, dr \,
r^{(d-1)/2} e^{-r^2/t} \Psi(r)$ where $\Psi(r)$ is a random field with
short range correlations. 
We have computed $Q(t)$ for $d=2, 10, 20, 30, 40$ and  $50$
and found, in all cases, $Q(t) \to Q_\infty$. 
The 
asymptotic values $Q_\infty$ as a function of $\theta$,
reported in Fig.~\ref{Fig2}c, are in good 
agreement, even quantitatively, with $Q^M(\theta)$ for multiplicative processes.
In Fig.~\ref{Fig2}c we have also reported the value of
$Q_\infty$ for another smooth process called the
random acceleration process, for which $\theta=1/4$. These data
suggest that smooth 
processes, at variance with coarsening dynamics in Ising systems, are better
approximated by multiplicative processes. 
%

A close look at 
Fig.~\ref{Fig1} suggests investigation of other closely related cousins
of  $l_{\rm max}(t)$ defined in Eq.~(\ref{def_lmax}), such as  $\mu_{\max}(t) = {\max} (\tau_1, 
\tau_2,\cdots, 
\tau_{N+1})$ or for instance $\lambda_{\max}(t) = {\max} (\tau_1, \tau_2,\cdots,
\tau_{N})$. For renewal processes with L\'evy index $\theta<1$, 
the analysis presented above can be extended to
obtain exact results for the average of both observables. In both cases, the
average  grows 
linearly with $t$ but with different $\theta$-dependent prefactors. While the prefactor
for the former case was computed in Ref.~\cite{scheffer} by a rather complicated but
rigorous method, the latter case $\langle 
\lambda_{\max}(t)\rangle$ has not been studied, to our knowledge, even for
Brownian motion ($\theta=1/2$). We find 
$\langle \lambda_{\max}(t) \rangle \simeq \lambda_{\infty}(\theta)\, t$
which yields, in particular, a new constant $\lambda_{\infty}(1/2) = 0.241749...$ for
Brownian motion. The detailed studies of 
$\mu_{\max}(t)$ and
$\lambda_{\max}(t)$ for other nonequilibrium processes will be
reported elsewhere~\cite{us_prep}.

In conclusion, we have shown that the average length of the longest excursion
has rather rich and universal asymptotic time dependence for a variety of
nonequilibrium processes. Our analytical and numerical results highlight
the importance of extreme value statistics in generic nonequilibrium dynamics.

\end{document}